# Securing Sideways

Thwarting Lateral Movement by Implementing Active Directory Tiering

Tyler Schroder[1] 0009-0008-1847-1712

Department of Computer Science, 51 Prospect St, New Haven CT 06511

The MITRE Corporation, 7515 Colshire Drive, McLean VA 22102, USA

tyler.schroder@yale.edu

Sohee Kim Park 0000-0002-8553-3468

Department of Computer Science, Yale University, 51 Prospect Street, New Haven CT 06511, USA,

sohee.park@yale.edu

The advancement of computing equipment and the advances in services over the Internet has allowed corporations, higher education, and many other organizations to pursue the shared computing network environment. A requirement for shared computing environments is a centralized identity system to authenticate and authorize user access. An organization's digital identity plane is a prime target for cyber threat actors. When compromised, identities can be exploited to steal credentials, create unauthorized accounts, and manipulate permissions—enabling attackers to gain control of the network and undermine its confidentiality, availability, and integrity. According to the FBI's Internet Crime Complaint Center (IC3), cybercrime losses reached a record of 16.6 B in the United States in 2024, a 33% increase from 2023. For organizations using Microsoft software, Active Directory is the on-premises identity system of choice. In this article, we examine the challenge of security compromises in Active Directory (AD) environments and present effective strategies to prevent credential theft and limit lateral movement by threat actors. These strategies include: 1) Active Directory Tiering—a method of logically segmenting the network into three tiers based on the sensitivity of the systems and data within each tier; 2) implementing the technical steps necessary to deploy AD Tiering; and 3) evaluating the benefits and potential trade-offs, such as setup time, hardware costs, ongoing maintenance, and required adjustments to business processes. Our proposed approaches aim to confine the movement of compromised credentials, preventing significant privilege escalation and theft. We argue that through our illustration of real-world scenarios, tiering can halt lateral movement and advanced cyber-attacks, thus reducing ransom escalation. Our work bridges a gap in existing literature by combining technical guidelines with theoretical arguments in support of tiering, positioning it as a vital component of modern cybersecurity strategy even though it cannot function in isolation. As the hardware advances and the cloud sourced services along with AI is advancing with unprecedented speed, we think it is important for security experts and the business to work together and start designing and developing software and frameworks to classify devices automatically and accurately within the tiered structure.

---

[1] Work performed while a student at Yale University. Approved for Public Release; Distribution Unlimited. Public Release Case Number 25-2090. The author's affiliation with The MITRE Corporation is provided for identification purposes only and is not intended to convey or imply MITRE's concurrence with, or support for, the positions, opinions, or viewpoints expressed by the author.

## 1 INTRODUCTION

In 2023, $12.5 billion was lost to cybercrime due to cybercrime threat actor activity [23]. A common thread in many of these incidents is the exploitation of Microsoft Active Directory (AD), which serves as the backbone of identity and access management in most Windows-based enterprise environments. Regardless of how the initial breach occurs—through phishing, supply chain compromise, or other means—threat actors frequently pivot laterally within a network to escalate privileges and access critical systems. Active Directory is a prime target in this process, as compromising AD credentials or exploiting its vulnerabilities can grant attackers broad control over an organization's digital infrastructure[22]. Ransomware operations provide a clear example of this pattern. While the initial infection often begins with a phishing email that tricks a user into executing malicious code on an endpoint such as a laptop or desktop [17, 24], the real damage occurs after the attacker leverages that foothold to move laterally and escalate privileges—often by abusing AD. The compromised user account may have limited privileges, but it can serve as a gateway for discovering and accessing more sensitive credentials or systems. Once sufficient access is obtained, the threat actor can deploy ransomware across the environment, encrypting critical data and demanding payment for decryption [7].

This type of attack is preventable or containable if organizations are willing to commit both technical and business resources. Organizations can take concrete steps to prevent this credential theft and lateral movement through Active Directory Tiering. Tiering is an intentional re-structuring of the AD environment to limit the scope of where an identity is valid, restricting or containing a threat actor's movement with stolen credentials. In a tiered AD setup, networked devices and identities are grouped into three risk classes (from highest to lowest risk): Tier 0, Tier 1, and Tier 2. Tier 0 contains identity and management systems, Tier 1 contains shared infrastructure such as file servers and application servers, and Tier 2 contains endpoints such as laptops, desktops, and mobile phones. Credentials are unique and should not be reused across tiers, mitigating opportunities for a threat actor to navigate within a network and cause harm.

In this article, we present a formal framework for tiered Active Directory security that integrates both practical implementation methodology and theoretical security principles. Unlike previous informal discussions, higher level overviews [7, 9, 11] or vendor-specific solutions [5][25] , we provide a comprehensive, vendor-neutral approach that organizations can implement using existing Active Directory infrastructure. This novel synthesis bridges the gap between theoretical security models and practical deployment, enabling organizations to enhance their Active Directory security posture without additional software investments. Our proposed solution reduces the risk of full-scale ransomware and other cyber threat compromises by blocking threat actors from moving laterally to access more critical systems and sensitive data.

This article is structured as follows: in Section 2, we discuss how tiering can logically isolate (or insulate) zones in a network. Section 3 provides a demonstration of how tiering stops lateral movement and advanced cyber-attacks on a network[2]. Section 4 discusses tiering implementation and considerations for organizations looking to implement it. Section 5 discusses related work. Section 6 concludes the article, and Section 7 contains possibilities for future work.

## 2 WHAT IS TIERING?

Microsoft Active Directory is a directory service in an organization using Microsoft products that manages authorized users, computers they are using, and permission they have in the environment [12]. It serves as a centralized database that keeps track of the information of the network of the organization. When a user logs into a computer with his credentials, AD checks if the user is a valid user, what resources (e.g., printers) should be allowed to the user, and what policies should be applied to the session. The password they enter is hashed and compared to the stored hash on the nearest domain controller.[3] If the hashes match, the user is granted access. AD defaults to allowing any user to log into any device in the network. This includes the crown jewel of the environment and a threat actor's end goal, the primary domain controller. A domain controller is a critically sensitive server that maintains an authoritative copy of the directory and handles authentication and authorization for the network. Threat actors take advantage of this default by compromising a regular user account and then move from device to device, checking each one for higher-privileged user accounts with the goal to break into a domain controller. For example, a helpdesk administrator may have logged into an end user's Windows laptop in to install software at a user's request. Windows will (by default) cache the administrator's credentials for a faster login experience and expose the credential in the system's cache. A threat actor can harvest the administrator's credentials from the local compromised system, can be authenticated as a higher-privileged account, and then gains access to new systems or data.

---

[2] Lateral movement is movement within a corporate network, where a user logs into another device from a previously authenticated one.

Although Tiering cannot completely stop threat actor movement—such a blocker would cripple network functionality, Tiering sets logical boundaries across AD, preventing large jumps in privilege to stop theft or destruction. An apt analogy is watertight bulkheads in a modern ship (see Figure 1 - Watertight Bulkheads (USNI)). If the hull is breached in one compartment (e.g., compartment A), water cannot spread to other compartments. Similarly, regardless of which tier a threat actor may compromise, the treat actor is limited to access only within that tier (e.g. a breach in Tier 1 or Tier 2 cannot allow access to higher tiers). A user with privileges to access only lower tiers cannot access (read up to) higher tiers. An attack in the opposite direction is possible when Tier 0, the highest tier, is directly compromised, as a threat actor can create their own identities to use in whichever tier they desire. Further, this opposite direction attack is possible (Tier 0 to Tier 1 or 2) because limited write down (access) from higher tiers into lower tiers is allowed to allow AD to function. For example, a management server may reach out from Tier 1 to manage Tier 2 devices in smaller networks. Domain Controllers in Tier 0 must be able to update credentials for devices in Tier 1 and 2. A tiering is illustrated in Figure 2 - Tiered Infrastructure Model.

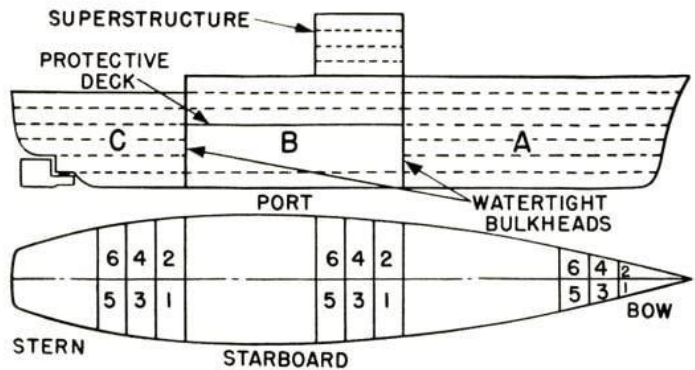

Figure 1 - Watertight Bulkheads (USNI) [8]

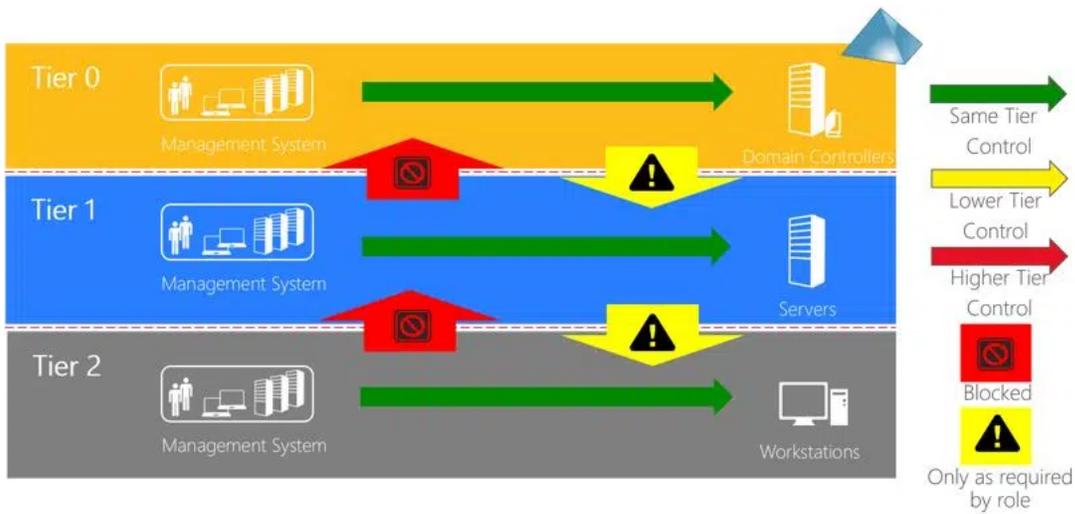

Figure 2 - Tiered Infrastructure Model [10]

In a tiered environment, there is no business requirement for a regular end-user account to log into the operating system of a file server or business app server. **A user with an identity in Tier 2 can only access devices within Tier 2.** Attempts to log into Tier 1 or 0 devices would be blocked by AD policy, and users would instead have multiple identities[4] in the network, scoped to each tier. An application owner who needs to manage servers in Tier 1 would have a separate identity, and an IT service provider would have a second additional separate identity for managing user accounts on the network in Tier 0. Table 1 - Access Needs and Accounts Summary offers a summary view, with the exception that some IT systems administrators may never need access to Tier 0 systems (helpdesk, software installation teams).

Table 1 - Access Needs and Accounts Summary

| User Type | Access Needs | Accounts |
|---|---|---|
| Regular Employee | Daily Workstation<br>Files on File Share | Tier 2 |
| Application Owner | Administrative rights over select Tier 1 servers<br>Daily Workstation | Tier 2<br>Tier 1 (limited to specific servers) |
| IT System Administrator | Administrative rights over Tier 0 systems<br>Administrative rights over Tier 1 systems<br>Administrative rights over Tier 2 systems<br>Daily Workstation | Tier 2<br>Tier 1<br>Tier 0 |

---

[4] Multiple identities require a unique username and password per tier. Identities may share the same 2FA device (Two-Factor Authentication Device, e.g. the same authenticator app on a phone) but should use unique secrets (unique one-time code generations).

## 3 HOW TIERING CONTAINS ATTACKS

In a typical AD compromise, threat actors gain a foothold in the corporate network via an end-user device such as a laptop or desktop. A threat actor then must move from this entry point to sensitive information. Under the tier model, such information would be in Tier 1. Lastly, threat actors create their own accounts to set up persistence on domain controllers that would be in Tier 0. Figure 3 – Sample attack map for a ransomware deployment illustrates an example network with an Active Directory server, some files and app servers, and a few endpoints. In our first network, there is logical tiering in place (yellow arrows). In our second network no such boundaries exist (purple arrows).

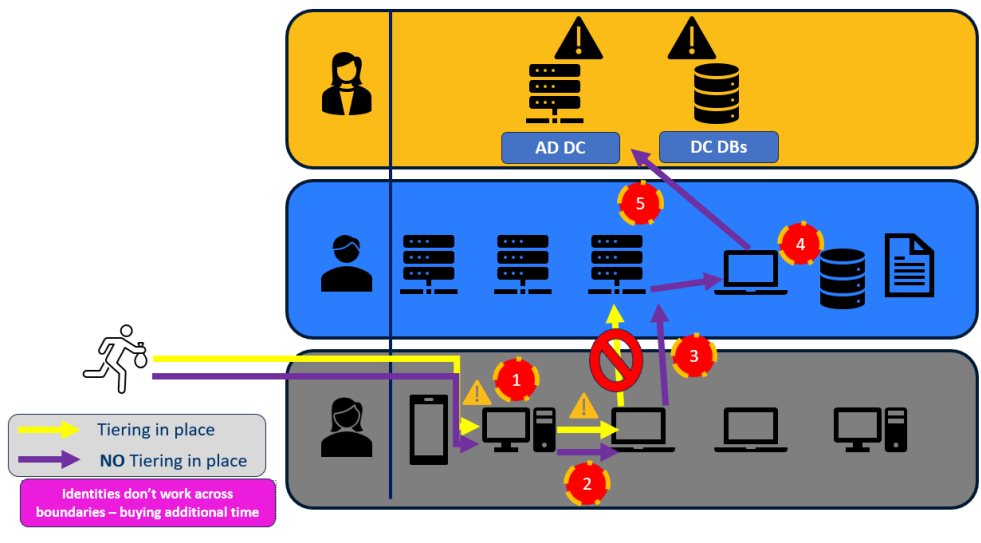

Figure 3 – Sample attack map for a ransomware deployment

In this attack, a threat actor is looking to gain complete dominance over a target network, and not to perform a standalone business email compromise or petty theft from a single user [3]. First, the threat actor will gain initial access into the organization's digital environment. A common approach is for a threat actor to use a phishing email to steal credentials or encourage the download of a remote access tool. However, threat actors may opt for other methods such as breaking into less-secured internet-of-things devices such as a smart thermostat [4, 6] or using social engineering to trick a helpdesk employee into resetting passwords [14]. Initial access is represented as **step 1** as illustrated in Figure 3.

After gaining access to a device with the initial stolen credentials, the threat actor will check to see what other accounts exist on the device (through the Local Security Authority Subsystem Service, or LSASS.exe) in hopes of finding an account with higher privileges than the current account. If their search is unsuccessful, the threat actor can log into other workstations as seen in **step 2** (where the threat actor moves from a desktop to a laptop). On the laptop, using the same original compromised account, the threat actor discovers an account called 'workstation admin' that has higher privileges. Using this new administrative account, the threat actor can move to higher-security devices that the company would consider confidential.[5] Threat actor can also see file servers the business relies on. The threat actor will then try to move from the laptop to the server in **step 3**.

---

[5] Data could range from trade secrets, intellectual property, accounts payable information, etc.

If the network is not tiered, AD grants a login to the threat actor with a stolen identity (that is valid anywhere). If the network is tiered, this attempt will be blocked because an identity cannot be used across tiers. AD does not allow a remote connection from a lower tiered device to a higher tier and blocks the use of a lower tiered identity on a higher tier device, which is not prevented in the regular network shown in **step 4**. If the network is not tiered, the threat actor can move from the file server to other devices and ultimately gain access to AD via the domain controller. Lastly, in **step 5**, the threat actor gains complete control over the AD and environment by moving from Tier 1 to Tier 0. If the network were tiered, the threat actor would never have left the Tier 2 device boundary. In the unlikely event the threat actor directly compromised a Tier 1 server and identity, they would still be blocked from accessing Tier 0 devices. Tiering contains the impact a threat actor can cause based on the point of entry into a network.

## 4   TIERING - A PART OF A GREATER CYBERSECURITY POSTURE

### 4.1   Integrating Tiering with Traditional Security Defenses

While tiering is a powerful security strategy, it is important to acknowledge that it does not replace traditional security methods such as two-factor authentication (2FA). Instead, tiering complements these methods by adding an additional layer of defense, making it much more difficult for threat actors to navigate within a network if they manage to compromise an account.

Two-factor authentication remains a critical element of cybersecurity. By requiring a second form of verification, it ensures that even if a threat actor gains access to a user's password, they will still need access to the second factor (such as a mobile device or security token) to authenticate. This significantly impedes unauthorized access. Instead, combining 2FA with a tiered security model provides enhanced protection. AD tiering segregates the network into distinct zones with separate credentials and permissions, mitigating the risk of credential theft and lateral movement:

1.  **Separate Identities**: For tiering to be effective alongside other security defenses, it is essential to have completely separate identities for each tier. This means that each user should have unique usernames, passwords, and second-factor secrets for each tier. This separation ensures that compromising credentials in one tier does not provide access to higher-security tiers.

2.  **Separate Second Factors**: each identity includes a separate 2FA secret. All tiered identities can use the same format of 2FA but would have different secrets for code generation. In the event a different identity's username, password, or 2FA secret is stolen, none of those pieces can be used to compromise other accounts.[6]

3.  **Enhanced Security Posture**: By implementing tiering, even if a threat actor bypasses 2FA and gains access to an account in Tier 2 (endpoints), they will still face significant barriers progressing to Tier 1 (shared infrastructure) or Tier 0 (identity and management systems). The logical separation reduces the chances of escalation and subsequent breaches.

4.  **Defense in Depth**: Tiering, coupled with strong authentication mechanisms like 2FA, exemplifies the principle of defense in depth. While 2FA secures the authentication process, tiering ensures that each layer of the network is independently secured, minimizing the impact of a compromised account.

---

[6] Ideally the 2FA method would be phishing-resistant FIDO or TOTP code generation. Use of email or SMS is not recommended per NIST SP 800-63. https://www.nist.gov/itl/smallbusinesscyber/guidance-topic/multi-factor-authentication

## 4.2 Implementing Tiering in an AD Environment

A successful tiering implementation is a four-part process: first, the AD environment must be secured; second, administrators must reduce excessive permissions in the environment; next, separate identities between tiers should be created; and finally, controls should deny access between tiers. Optionally, privileged access workstations can be rolled out to support tier 0 management.

AD tiering cannot work alone - it relies on a secure AD baseline. Prior to implementing tiering, organizations should review relevant benchmarks (such as the Center for Internet Security (CIS) AD benchmarks) to secure their AD environment.

With a secure baseline, the next step is to identify high risk permissions and remove or reduce them wherever possible. In AD, the highest-privileged (and most risky) roles are Domain Administrator, Enterprise Administrator, and Schema Administrator.

With a proper security posture in place, organizations should evaluate their existing infrastructure and classify into the three tiers. Tier 0 contains the AD Domain Controllers, and any systems or identities used to manage them. Tier 1 contains shared business infrastructure, and Tier 2 contains user-facing assets. Below, Table 2 lays out in which tiers typical network devices would belong. Tables 3 through 5 discuss finer details for each tier.

Table 2 - Tier membership suggestions

| Tier 0 | Tier 1 | Tier 2 |
|---|---|---|
| Domain Controllers<br>Systems used to manage Domain Controllers | Application Servers<br>File Servers<br>Shared infrastructure used by multiple users | Printers, Phones, Virtual Desktops, mobile devices<br>Kiosks or public facing infrastructure<br>Higher risk devices |
| Identities with permissions to log into or administer tier 0 systems | Identities with permissions to log into or administer tier 1 systems | Identities with the ability to log into or administer tier 2 systems |

Tiers should be created in AD as separate high-level organizational units. User and computer objects should be moved from shared or default storage locations into their proper tier organizational unit (OU). Beyond where accounts live, it's important to review unique group and user permissions to ensure that an identity cannot take action on accounts in other tiers. Specific AD services may also operate at higher permissions and should be classified into a higher tier. Tables 4, 5, and 6 below walk through each tier and considerations for each

Table 3 - Tier 0 Principals, Controls, and Services

| Principals | Control | AD Services |
|---|---|---|
| Users | Take Ownership | Domain Controllers |
| Security Groups | Change ACLs | Privileged Groups |
| Servers | Read Secrets | Configuration or Settings |
| Endpoints managing Tier 0 | Full Write | NTDS |

Table 4 - Tier 1 Principals, Controls, and Services

| Principals | Control | AD Services |
|---|---|---|
| Application or Tier 1 Server Administrators<br>Individual Application Administrator accounts on shared Tier 1 hardware | Any member of a tier 1 group<br>Permissions equivalent to any tier 1 group<br>Administrative rights over a tier 1 application | Hardware running tier 1 services<br>Devices with cached, stored, or uses Tier 1 credentials |

Table 5 - Tier 2 Principals, Controls, and Resources

| Principals | Control | Resources |
|---|---|---|
| End users<br>Computer accounts | Any member of a tier 2 group<br>Permissions equivalent to any tier 2 group<br>Administrative rights over a tier 2 application<br>Helpdesk or Endpoint administrators | Workstations<br>Laptops<br>Mobile Devices<br>Tablets<br>High-risk or publicly exposed assets |

After assets are divided, additional user accounts should be provisioned and placed into the proper tiers. After accounts are created, administrators should deny access for lower tiers into higher tiers. This relies on a group policy object (GPO) to set logon permissions across each tier. One GPO per tier is applied to allow only the relevant tier to log in and deny access to the others.

### 4.3 Costs and Benefits

Tiering allows an organization to contain a cyberattack within a specific tier of infrastructure and minimize the risk of threat actors moving throughout their environment. However, like any technical solution, tiering is not without drawbacks. Setup time, hardware cost, maintenance, and business process adaptation are the core four areas to consider in implementing the proposed solutions.

**Setup time:** Systems administrators must review all groups and users for rogue permissions and unneeded administrative access to prevent jumping between tiers. This includes identifying users accessing files or network entities beyond their defined roles, or devices connecting to the organization's network without authentication. Rogue permissions expose security vulnerabilities and can invite cyber attacker's attempt for malicious activities. Organizations must provision up to two more accounts per user[7], depending on their roles and responsibilities, and end users must know how and at which tiers they should use these identities.

**Hardware costs:** An administrator may need multiple machines to prevent logging in with higher tier accounts on lower tier devices. Organizations must evaluate if they have budget and staff time to issue extra devices and support additional hardware. Some access could be provided through virtual machines, but existing hardware must be able to support the added load.

**Ongoing maintenance**: Care must be taken as new users, devices, and software are introduced into the network environment. Permissions and tiers must be scoped correctly to prevent unintended vulnerability or exposure in the tiers.

**Business processes:** Business applications and processes must be tested to ensure they work under the new tiered model. Legacy software may expect administrative permissions or unfettered access to a network to perform business functions. The average case end user will experience no impact in their day-to-day operation. Users with IT management duties will need to adapt to using a different identity and device for management tasks in different tiers.

---

[7] Consisting of a unique username, password, and 2FA secret per tier & identity

Table 6 – Tiering Benefits and Costs of Tiering Summary

| Benefits | Costs |
|---|---|
| Reduce lateral movement opportunities. | Set up time and effort. |
| Reduce ransomware breakout risk. | Hardware costs. |
| Reduce cached credential harvesting opportunities. | Ongoing maintenance to ensure separation of tiers. |
| | Expanded business processes to respect new logical divisions in network. |

## 5  RELATED WORK

Work on tiered access came as a part of the greater industry shift from perimeter-based security to zero-trust security[1, 16]8   Some academic studies exist, such as Erhan Sindiren's application security model [21] and earlier work in distributed systems security by Jean Bacon [2]. Microsoft owns the most recent work in the AD security world, which is understandable as they own the Active Directory and Entra ID products [18]. Documentation from Microsoft is available online for information technology professionals looking to switch their business networks to tiered postures [10, 13, 19, 20]. Some private technology consulting and security firms also  implement tiered postures as a service [15]. However, none of these resources consider both the business and technology case for tiered access in the same document, a gap that this work fills.

## 6  CONCLUSIONS

The accelerating pace of cybercrime—and increasing ease for threat actors to compromise networks—underscores the urgent need for organizations to secure their digital identity infrastructure. Microsoft Active Directory (AD), which serves as the dominant on-premises identity and access management system, presents a high-value target for threat actors due to its central role in authenticating users and authorizing access across enterprise networks. This work proposes a technical segmentation solution for the problem of credential-based attacks and lateral movement within AD environments, which remain central vectors for ransomware deployment and advanced persistent threats.

To address the threat, we proposed Active Directory Tiering as a practical and scalable defense-in-depth strategy. By segmenting systems into logically defined tiers based on risk and privilege level, AD Tiering constrains the ability of adversaries to move laterally across a compromised environment. This work bridges a gap in the literature by synthesizing theoretical concepts of identity segmentation with practical deployment guidance rooted in real-world constraints. While AD Tiering is not a standalone solution and must be complemented by modern endpoint detection and identity protection tools, use of tiering can significantly reduce the blast radius of credential theft and impedes privilege escalation when properly implemented.

## 7  FUTURE WORK

This work is primarily focused on the concerns and challenges around implementation of AD tiering. Extension works could provide deployment scripts in Microsoft PowerShell (or other configuration management software like Ansible for Puppet) to help companies implement tiering. Future studies could also create frameworks for how to classify devices as Tier 0, 1, or 2 to assist companies with the tiering process.

---

8 Perimeter security was the concept of a hardened external perimeter with an implicitly trusted internal network. Zero-trust is the shift to authentication and authorization at every stage, with no such implicit internal trust.